\documentstyle[aps,prd,twocolumn,epsf]{revtex}

\newcommand{\be}{\begin{equation}}
\newcommand{\ee}{\end{equation}}
\def\reff#1{(\ref{#1})}
\def\spose#1{\hbox to 0pt{#1\hss}}
\def\ltapprox{\mathrel{\spose{\lower 3pt\hbox{$\mathchar"218$}}
 \raise 2.0pt\hbox{$\mathchar"13C$}}}
\def\gtapprox{\mathrel{\spose{\lower 3pt\hbox{$\mathchar"218$}}
 \raise 2.0pt\hbox{$\mathchar"13E$}}}

\begin{document}
\draft
 

\title{$SU(2)$ Landau gluon propagator on a $140^3$ lattice}
 
\author{Attilio Cucchieri\thanks{
 Email:~{\tt attilio@if.sc.usp.br},~{\tt mendes@if.sc.usp.br}~and
       {\tt taurines@if.ufrgs.br}} and
       Tereza Mendes$^{*}$}
\address{{\small Instituto de F\'\i sica de S\~ao Carlos,
                 Universidade de S\~ao Paulo \\
                 C.P.\ 369, 13560-970 S\~ao Carlos, SP, Brazil}}
\author{Andre R.\ Taurines$^{*}$}
\address{{\small Instituto de F\'\i sica,
                 Universidade Federal do Rio Grande do Sul \\
                 Av.\ Bento Gon\c calves, 9500, Campus do Vale \\
                 91501-970 Porto Alegre, RS, Brazil \\[3mm]}}

\date{\today}

\maketitle

\begin{abstract}
We present a numerical study of the gluon propagator in
lattice Landau gauge for three-dimensional pure-$SU(2)$
lattice gauge theory at couplings $\beta = 4.2, 5.0, 6.0$ and for
lattice volumes $V = 40^3,\, 80^3, \,140^3$. In the limit
of large $V$ we observe a decreasing
gluon propagator for momenta smaller than
$p_{dec} = 350^{+ 100}_{- 50} \, \mbox{MeV} $.
Data are well fitted by Gribov-like formulae
and seem to indicate an infra-red critical exponent $\kappa$
slightly above 0.6, in agreement
with recent analytic results.
\end{abstract}

%
%


\section{Introduction}

The study of the infra-red (IR) limit of QCD is of central
importance for understanding the mechanism of quark
confinement and the dynamics of partons at low energy.
Despite being non-gauge-invariant, the gluon
propagator is a powerful tool in this
(non-perturbative) investigation \cite{DSE}.
In particular, it would be interesting to express it
in a closed form for recovering the
phenomenology of Pomeron exchange from first principles
\cite{Pomeron}.

Studies of the coupled set of Dyson-Schwinger equations for
gluon and ghost propagators in Landau gauge predict for the
gluon propagator an IR behavior of the form $D(p) \sim
p^{4 \kappa - 2}$ [implying $D(0) = 0$ if $\kappa > 0.5$].
The available predictions for the IR exponent are 
$\kappa \in [0.52, 1.00]$ in the four-dimensional case
\cite{IRkappa,Zwanziger:2001kw}
and $\kappa \approx 0.648$ or $\kappa = 0.75$ in three dimensions
\cite{Zwanziger:2001kw}.

Furthermore, in the minimal Landau gauge, the gauge-fixed
configurations belong to the region $\,\Omega\,$ of
transverse configurations, for which the Faddeev-Popov operator
is non-negative. This implies a rigorous inequality
\cite{vanishing}
for the Fourier components of the gluon field $ A_{\mu}(x) $
and a strong suppression of the gluon propagator in the IR limit.
In particular, for dimension $d$ and infinite volume,
it is proven that the (unrenormalized) gluon propagator
is less singular than $p^{2-d}$ and that,
very likely, it vanishes in the IR limit
\cite{vanishing}.
A vanishing gluon propagator at $p = 0$, given by the
form $ p^{2} / ( p^{4} \,+\, \lambda^4) $, was also obtained
by Gribov
\cite{Gribov:1977wm}. 
Here the mass scale $\lambda$ arises
when the configuration space is restricted to the region $\Omega$.
A generalization of this expression has been introduced in Ref.\
\cite{Stingl:hx}
as an Ansatz for a non-perturbative solution
of the gluon Dyson-Schwinger equation.


Numerical studies
\cite{Cucchieri:1999sz,Bonnet:2001uh}
have now established that the gluon propagator in
lattice Landau gauge shows a turnover in the IR
region and attains a finite value for $p = 0$.
Evidences of a decreasing gluon propagator for small $p$
have been obtained in the $4d$ $SU(2)$ and $SU(3)$ cases
(but only in the strong-coupling regime)
\cite{Cucchieri:1997dx,4dIR}
in the $3d$ $SU(2)$ case (also in the scaling region)
\cite{Cucchieri:1999sz,Cucchieri:2000cy,Cucchieri:2001tw},
in the $3d$ $SU(2)$ adjoint Higgs model
\cite{Cucchieri:2001tw},
in the $4d$ $SU(2)$ case at finite temperature
\cite{Bogolubsky:2002ui}
and for the equal-time three-dimensional transverse
gluon propagator in $4d$ $SU(2)$ Coulomb gauge
\cite{Cucchieri:2000kw}.
In this last case, one obtains an excellent fit of the
transverse propagator by a Gribov-like formula.

This work aims to verify the possibility
of using Gribov-like formulae to fit data of the
gluon propagator also in Landau gauge. At the same time, we
will try to obtain a value for the IR critical exponent
$\kappa$ to be compared to the analytic determinations mentioned
above.
In order to probe the infinite-volume limit and the IR region we
consider the three-dimensional case and the $SU(2)$ group,
using lattice sizes up to $140^3$. Note that the study of the
gluon propagator in three dimensions is also of interest in
finite-temperature QCD
\cite{HTQCD}.


\section{Numerical simulations}
\label{sec:numeri}

We consider the standard Wilson action for $SU(2)$ lattice
gauge theory in three dimensions with periodic boundary
conditions. The numerical code is entirely parallelized using
MPI. (Technical details and a study of the code performance
are left for a subsequent work
\cite{Gonzalo}.)
For the construction of staples we follow Ref.\
\cite{HiokiPC}.
For the random number generator we use a double-precision
implementation of RANLUX (version 2.1) with luxury level set to 2.
Computations were performed on the PC cluster at the IFSC-USP.
The system has 16 nodes and a server with 866 MHz Pentium III CPU
and 256/512 MB RAM memory. The machines are connected with a
100 Mbps full-duplex network. The total computer time used for the
runs was about 80 days on the full cluster.


In Table \ref{tab:runs} we report, for each coupling $\beta$ and
lattice volume $V$, the parameters used for the simulations.
All our runs start with a random gauge configuration. For
thermalization we use a {\em hybrid overrelaxed} (HOR) algorithm
\cite{HOR}.
Each HOR iteration
consists of one heat-bath sweep over the lattice followed by $m$
micro-canonical sweeps. We did not try to find the best tuning for
$m$; we use $m = 4$ for $V = 40^3,\, 80^3$ and
$m = 5$ for $V = 140^3$. In order to optimize the
heat-bath code, we implement two different $SU(2)$
generators, namely methods 1 and 2 described in
\cite[Appendix A]{Edwards:1991eg}
with $h_{cutoff} = 2$.

 
For the numerical gauge fixing we use the {\em stochastic
overrelaxation} (SOR) algorithm
\cite{gfix,Cucchieri:2003fb}
with even/odd update. In Table \ref{tab:runs} we
report the value of the tuning parameter $p_{sor}$ used for each pair
$(\beta\mbox{,} \,V)$. We stop the gauge
fixing when the average value of $[ (\nabla \cdot A)^b(x) ]^{2}$
is smaller than $10^{- 12}$.
(For a definition of the lattice gauge
field $ A^b_{\mu}(x) $ and of the lattice divergence $ \nabla $
we refer to
\cite{Cucchieri:1999sz}.)
We note that half of the configurations for $V = 80^3$
were done using the so-called {\em Cornell method}
\cite{gfix,Cucchieri:2003fb},
with tuning parameters $\alpha_{corn} = 0.325, 0.32, 0.316$
respectively at $\beta = 4.2, 5.0, 6.0$.
In fact, the Cornell method is
somewhat faster than the SOR algorithm and leads to a
gauge fixing of comparable quality if one uses an even/odd update
\cite{Cucchieri:2003fb}.
A good estimator of the quality of the gauge fixing is the
quantity $\Sigma_Q$ (see eq.\ 6.8 in Ref.\
\cite{Cucchieri:2003fb}),
which should be zero when the configuration is gauge-fixed.
By averaging over the gauge-fixed configurations,
we find (at $\beta = 4.2$) that the ratio between the final
and the initial values of $\Sigma_Q $ is (in $95 \%$ of the cases) about
$5.3 \times 10^{-10}$ with the Cornell method and
$2.4 \times 10^{-11}$ with the SOR algorithm.
At the same time, the average CPU-time needed for updating each
site variable is about $11 \%$ smaller for the Cornell method.
In any case, the CPU-time for gauge fixing was quite significant for the
large lattices. In order to go to even larger lattices
one should probably implement a global algorithm such as the Fourier
acceleration method
\cite{gfix,Cucchieri:2003fb}
with the multigrid or conjugate gradient implementations introduced
in Ref.\
\cite{Cucchieri:1998ew},
which are highly parallelizable.

 
We ran the $40^3$ lattices on a single node, the $80^3$ lattices on
two nodes and the $140^3$ lattices on four nodes.
The parallelization of the code worked well. In fact, for runs on
2 and 4 nodes, we obtain speed-up factors 1.82 and 3.41 for the
heat-bath link update. For the micro-canonical link update the
factors are respectively 1.87 and 3.77 and for the SOR
site update we get 1.90 and 3.72.


For each $\beta$ we evaluate the average plaquette
$ \langle W_{1,1} \rangle $ (see Table \ref{tab:volu}). Results are
in agreement with the data reported in Ref.\
\cite{Cucchieri:1999sz},
but we now have smaller statistical errors.
(The data have been analyzed using various methods,
described in footnote 4 of Ref.\
\cite{Cucchieri:1999sz}.
Here we always report the largest error found.)
We also evaluate the tadpole-improved coupling
$\beta_{I} \equiv \beta\,
\langle W_{1,1} \rangle $. In this way, by using the fit given in
eq.\ 2 and Table IV of Ref.\
\cite{Lucini:2002wg}
we calculate the string tension $ \sqrt{\sigma} $ in lattice units
(see Table \ref{tab:volu}) and the inverse lattice spacing $a^{-1}$
using the input value $\sqrt{\sigma} = 0.44$ GeV.
The fit is valid for $\beta \gtapprox 3.0$,
i.e.\ the couplings $\beta$ considered here are well
above the strong-coupling region.
Let us notice that, if we compare the data for the string tension
(in lattice units) with data obtained for the $SU(2)$ group in four
dimensions \cite[Table 3]{Fingberg:1992ju},
then our values of $\beta$ correspond to $\beta \approx 2.28, 2.345,
2.41$ in the four dimensional case.
Finally, in  the same table we report the lattice spacing $a$ in fm
and the smallest non-zero momentum (in MeV) that can be considered
for each $ \beta$.
Thus, with the lattice volumes and the $ \beta $ values
used here we are able to consider momenta as small as 59 MeV
(in the deep
IR region) and physical lattice sides almost as large as 25 fm.


In this work we did not do a systematic study of Gribov-copy
effects for the gluon propagator. However, we
compared data obtained using the SOR and the Cornell
gauge-fixing methods. In principle, different gauge-fixing
algorithms --- or even the same algorithm with different
values of the tuning parameter
\cite{Giusti:2001xf} ---
can generate different Gribov copies starting from the same
thermalized configuration. Thus, this comparison provides an
estimate of the bias (Gribov noise) introduced by the gauge-fixing
procedure. We found that, in most cases, the
difference between the two sets of gluon-propagator data is within
1 standard deviation and that, in all cases, it is
smaller than 2 standard deviations. Moreover, this difference
did not show any systematic effect, suggesting that
the influence of Gribov copies on the gluon propagator (if present)
is of the order of magnitude of the numerical accuracy. This is 
in agreement with previous studies in Landau gauge for the $SU(2)$
and $SU(3)$ groups in four dimensions 
\cite{Cucchieri:1997dx,Giusti:2001xf,Mandula:nj}.
A similar result has also been obtained for the
Coulomb gauge
\cite{Cucchieri:2000gu}.
Note that, in the $U(1)$ case
\cite{U1},
Gribov copies can affect the
behavior of the photon propagator, making it difficult to
reproduce the known perturbative behavior in the Coulomb phase.
The situation is very different for
the $SU(2)$ and $SU(3)$ cases, at least when considering
the lattice Landau gauge.
In fact, as said in the Introduction,
in the non-Abelian case the minimal-Landau-gauge condition
implies 
\cite{vanishing}
the positiveness of
the Faddeev-Popov matrix and a strict bound for
the Fourier components of the gluon field $A_{\mu}(k)$.
The bound applies to all Gribov copies obtained with the numerical
gauge fixing. Thus, if the behavior of the gluon propagator
is determined by this bound 
\cite{vanishing,Gribov:1977wm},
then this behavior should be the same for all lattice Gribov copies.
This would explain why we do not see Gribov-copy effects here.
Clearly, the same result does not apply to the
$U(1)$ theory, since in this case the Faddeev-Popov matrix is independent
of the gauge field.

 
\section{Results and Conclusions}

We evaluate the lattice gluon propagator $D(k)$ and study it
as a function of the lattice momentum $ p^2(k)$
(see Ref.\
\cite{Cucchieri:1999sz}
for definitions).
In our simulations we consider, for each gauge-fixed 
configuration, all vectors $k \equiv (k_{x}\,, k_{y}\,, k_{t})$
with only one component different from zero and
average over the three directions.
For the gluon propagator we analyze the data by
estimating the statistical error with three different
methods: standard deviation, jack-knife with single-data
elimination and bootstrap (with 10000 samples).
We found that the results obtained are in agreement
in all cases. Here we always use the standard-deviation
error.

Let us recall that in the three-dimensional case
the coupling $g^2$ has dimensions of mass.
Thus, in order to obtain a dimensionless
lattice coupling we have to set $\beta = 4/(a g^2)$.
Then, with our notation \cite{Cucchieri:1999sz},
the quantity $\,a \,D(k)\,$ approaches $\, g^2 D^{(cont)}(k)/ 4 $
in the continuum limit,
where $\,D^{(cont)}(k)\,$ is the unrenormalized continuum gluon
propagator.

\begin{figure}[t]
\begin{center}
\vspace*{-4cm}
\epsfxsize=0.40\textwidth
\leavevmode\epsffile{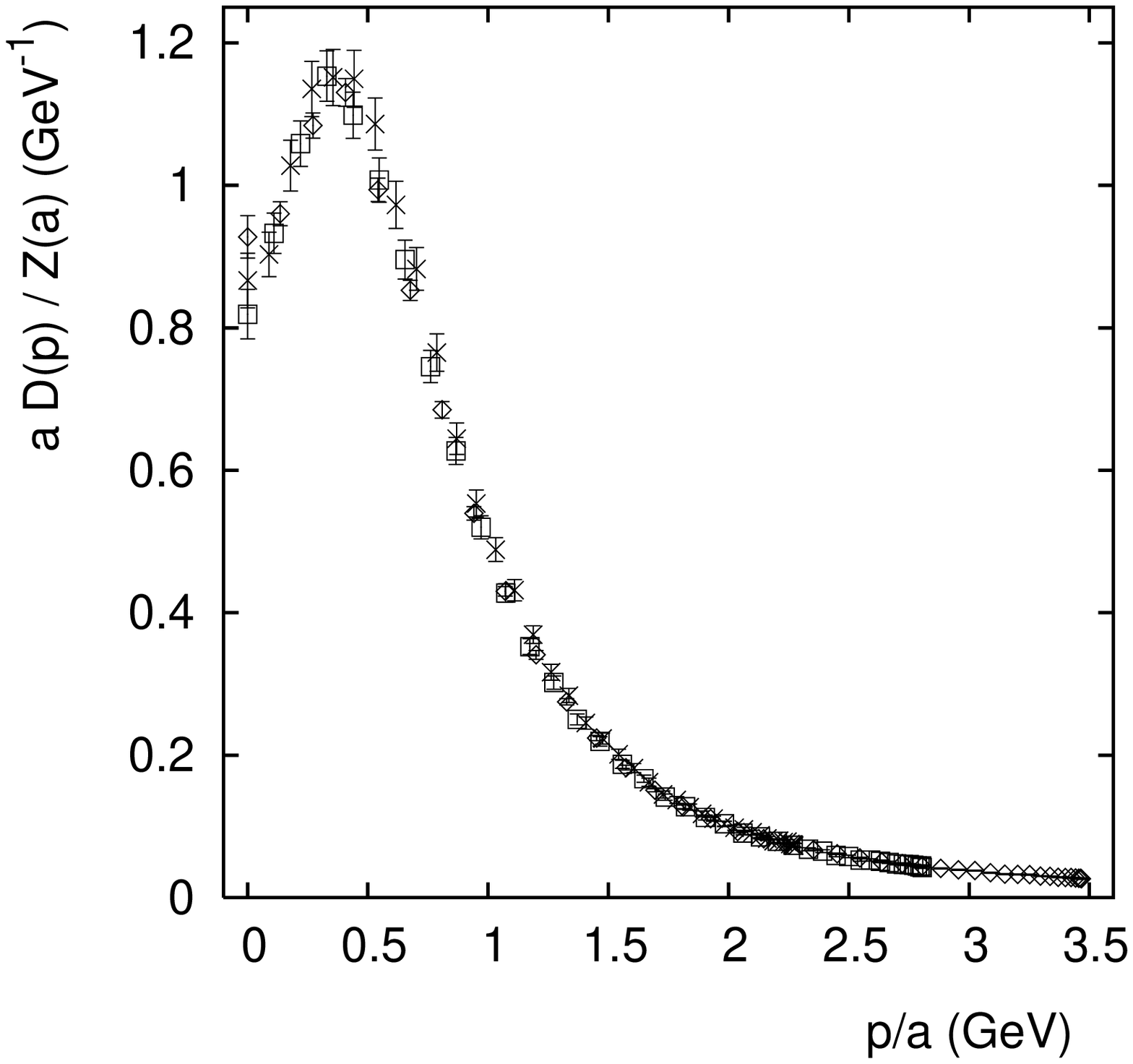}
\hspace*{1.4cm}
\epsfxsize=0.40\textwidth
\leavevmode\epsffile{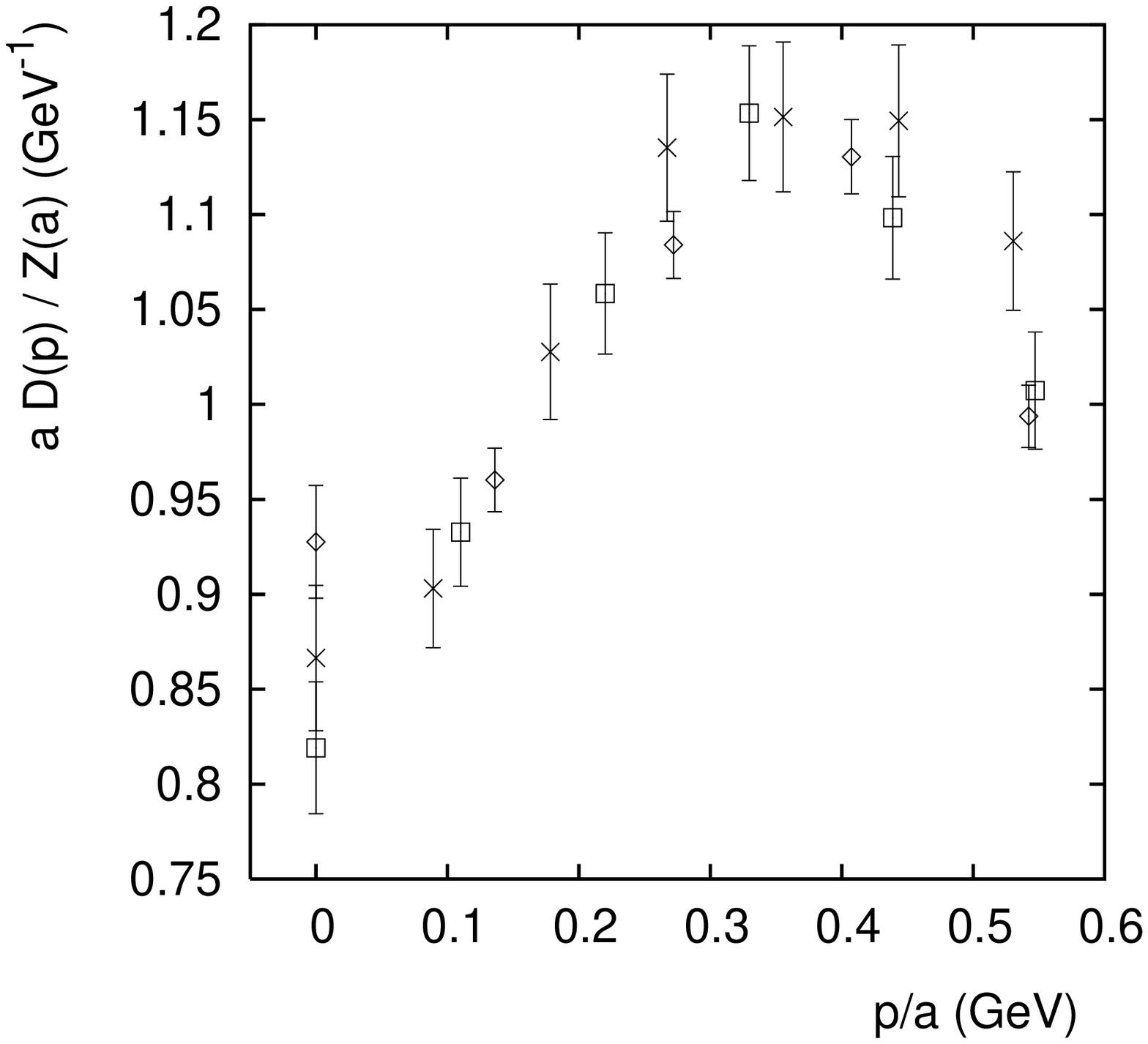} \\
\end{center}
\caption{~Plot of the rescaled gluon propagator as
a function of the lattice momentum for
$V=80^3$ and $\beta=4.2\,(\times)$, $\,5.0\,(\Box)$, $6.0\,(\Diamond)$.
The second plot shows only the IR region.
Error bars are obtained from propagation of errors.
}
\label{fig:fits}
\end{figure}

In order to compare lattice data at different $\beta$'s, we
apply the matching technique described in 
\cite[Sec.\ V.B.2]{Leinweber:1998uu}.
(Note that we have already determined the lattice spacing $a$,
as described above.)
We start by checking for finite-size effects, comparing data at
different lattice sizes and same $\beta$ value. In this way,
we find (for each $\beta$) a range of ultra-violet (UV) momenta for which
the data are free from finite-volume corrections. We then
perform the matching using data for these momenta and
$V = 40^3$, since for this lattice volume the
errors are smallest (about $1 \%$).
In particular, when matching data obtained at two
different values of $\beta$, we first interpolate the data
for the larger $\beta$ (the fine lattice) using a spline.
Finally, we find the multiplicative factor $R_Z = Z(a_f) / Z(a_c)$
corresponding to the best fit of the (multiplied) coarse-lattice
data to the same spline. The error of $R_Z$ is estimated using
a procedure similar to the one described in
\cite{Leinweber:1998uu}.
The method works very well (see Figure \ref{fig:fits}). 
Notice that we did not
fix the remaining global factor $Z$ imposing a
renormalization condition, as done for example in 
Ref.\
\cite{Alexandrou:2002gs}.
Our case is equivalent to
setting $Z(a) = 1$ at $\beta = 6.0$.

The data obtained after the matching are shown for $V = 80^3$
in Figure \ref{fig:fits}.
Clearly, we find that the gluon propagator decreases in
the IR limit for momenta smaller than $p_{dec}$, which
corresponds to the mass scale $\lambda$ in a Gribov-like
propagator.
From the plot we can estimate $p_{dec} = 350^{+ 100}_{- 50} \,
\mbox{MeV}$, in agreement with Ref.\
\cite{Cucchieri:1999sz}.

\begin{figure}[b]
\begin{center}
\vspace*{-3.6cm}
\epsfxsize=0.40\textwidth
\leavevmode\epsffile{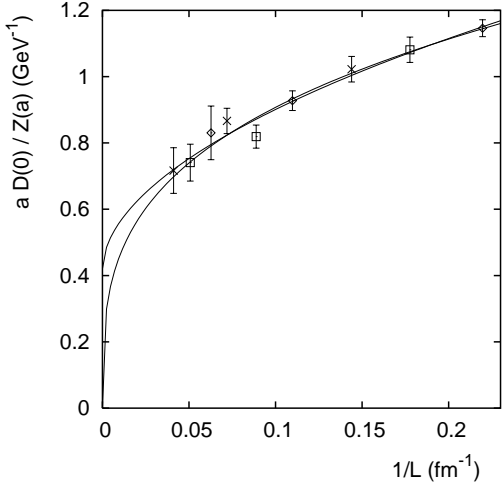}
\end{center}
\caption{~Plot of the rescaled gluon propagator
at zero momentum as a function of the inverse lattice side
for $\beta=4.2\,(\times)$, $\,5.0\,(\Box)$, $6.0\,(\Diamond)$.
We also show the fit of the data using the Ansatz
$d + b / L^c$ both with $d = 0$ and $d \neq 0$.
Error bars are obtained from propagation of errors.
}
\label{fig:zero}
\end{figure}

In Figure \ref{fig:zero} we plot the rescaled gluon
propagator at zero momentum, namely $a D(0) / Z(a)$, as a
function of the inverse lattice side $L^{-1} = 1/(a N)$ in physical
units (fm$^{-1}$). We see that $a D(0) / Z(a)$
decreases monotonically as $L$ increases, in agreement with
Ref.\
\cite{Bonnet:2001uh}.
It is interesting
to notice that these data can be well fitted using the
simple Ansatz $d + b / L^c$ both with $d=0$ and $d \neq 0$
(see Figure \ref{fig:zero}). In order to decide for one or the
other result one should go to significantly larger lattice sizes.
We plan \cite{3dprep} to extend
these simulations to $\beta = 3.4$
and lattice sizes up to $260^3$, allowing us to
consider a value $L^{-1} \approx 0.017$ fm$^{-1}$.
(This requires running in parallel on
all nodes of our PC cluster.)


Following Ref.\
\cite{Cucchieri:2000kw}
we fit
the data using a Gribov-like (or Stingl-like) formula
\begin{equation}
D(p) = \frac{s + z\, p^{2\alpha}}{y^2 + (p^2 + x)^2}\,\mbox{,}
\label{eq:gribov}
\end{equation}
where $z, s, \alpha, x$ and $y$ are fitting parameters.
For non-negative $\alpha$ this implies a finite gluon propagator
in the IR limit, with a behavior given by
$D(p) \propto (s + z\, p^{2\alpha})$.
If $y^2 > 0$ this form corresponds to a propagator with poles at
$m_{\pm}^2 = -x \pm i y$, while if $y^2 \leq 0$ the poles are real.
Finally, note that the IR exponent $\kappa$ considered in
studies of Dyson-Schwinger equations is given in terms of $\alpha$
by $\kappa = (1 + \alpha)/2$, assuming $D(0) = 0$.
Results of our fits using the un-rescaled lattice data $D(k)$
are reported in Table \ref{tab:par}. We see
that $\alpha$ decreases when the physical lattice volume increases.
Also, we always get $y^2 > 0$ and $x < \left|y\right|$,
which seems to support the scenario of purely imaginary poles
found in \cite{Cucchieri:2000kw}.
Let us recall that the poles
of the gluon propagator are gauge-invariant (at all orders
in perturbation theory)
\cite{gdepen}.

Notice that this fitting form leads in general
to the wrong UV
behavior, namely $D(p) \sim p^{2\alpha-4}$. This is not
a serious problem, since the largest momentum we can consider is
about $3.5$ GeV, i.e.\ we are not really exploring the UV
limit. On the other hand, the exponent $\alpha$ in \reff{eq:gribov}
plays a role both in the IR and in the UV regimes. Thus,
the values obtained for $\alpha$ probably correspond to
averaging the behavior of the gluon propagator in these two regions.
In particular, since we expect $D(p) \sim p^{-2}$ in the UV limit,
it is likely that the IR behavior of the propagator
be given by a smaller exponent $\alpha$ than the ones
reported in Table \ref{tab:par}. To check this we set $x=0$
and introduce an anomalous dimension $\gamma$:
\begin{equation}
D(p) = \frac{s + z\, p^{2 \alpha}}{y^2 + p^{2(1 + \gamma)}}\,\mbox{.}
\label{eq:newgribov}
\end{equation}
Results for this fitting form are also reported in Table \ref{tab:par}.
Indeed, we get values of $\alpha$ smaller than in the previous
case (and still decreasing with increasing physical lattice volume).
For the anomalous dimension we obtain $\gamma \approx 0.65$, with
small volume- and $\beta$-dependence.
The problem with eq.\ \reff{eq:newgribov} is that
the introduction of $\gamma$ compromises the pole
interpretation.

Finally, one can also try the form
\begin{equation}
D(p) = \frac{s + z\, p^{2 \alpha}}{(y^2 + p^4)^{\gamma}}\,
\mbox{.}
\label{eq:newgribov2}
\end{equation}
This corresponds to a propagator with purely
imaginary poles $m_{\pm}^2 = \pm i y$ and at the same time
allows the data to select the IR and the UV behaviors
separately. The problem in this case is that the fit is
unstable for small lattice volumes. On the contrary, for $V = 140^3$
the fit works well and we obtain
$\alpha = 0.27(6)$, $0.29(7)$, $0.38(8)$
respectively for $\beta = 4.2, 5.0, 6.0$ and $\gamma \approx 0.72$.
Hence, our $\alpha$ values are of the order of $0.3$, corresponding  
to $\kappa \approx 0.65$. (Again,
there is a decrease of $\alpha$ when the physical lattice volume
increases.)


In order to check for possible effects from
the breaking of rotational invariance
\cite{Becirevic:1999uc} we redid our fits
substituting $ p^2(k) $ by $ \tilde{p}^2(k)
\equiv p^2(k) + p^{[4]}(k) / 12 $ (see Ref.\ \cite{Ma:1999kn}).
One expects this modification to play an important role in
the UV limit and to have a small effect in the IR case.
For all $\beta$ values and lattice volumes we obtain good
fits and results similar
to those reported above. In particular, using eq.\ 
\reff{eq:gribov} one still finds $y^2 > 0$ and $x < 
\left|y\right|$, supporting the scenario of purely imaginary
poles. At the same time, for the three fitting functions,
$ \alpha $ decreases when the physical lattice volume increases,
but its value is about $20-30\%$ larger than the one reported
in Table \reff{tab:par} and above. In particular, using
eq.\ \reff{eq:newgribov2} and data for $ V = 140^3 $ we
obtain $\alpha = 0.32(7)$, $0.39(8)$, $0.59(12)$
respectively for $\beta = 4.2, 5.0, 6.0$ and $\gamma \approx 0.7$.
This implies slightly
larger values for $\kappa = (1 + \alpha) /2 $. The analysis of
these discretization effects and their influence on the
extrapolation of $\alpha$ and $\kappa$ to the continuum limit
will be done elsewhere \cite{3dprep}. 

\vskip 2mm

We have confirmed, by numerical simulations in the
scaling region, that the transverse gluon propagator
in $3d$ $SU(2)$ Landau gauge is a decreasing
function for momenta $p \ltapprox 350 \, \mbox{MeV} $
(attaining a finite value at $p = 0$).
Also, the data are well fitted by the Gribov-like
formulae \reff{eq:gribov}--\reff{eq:newgribov2}
and we obtain an IR critical exponent $\kappa$
in agreement with recent analytic results.
In order to eliminate remaining discretization
and finite-volume effects
[and in particular to check if $D(0) = 0$],
we need to simulate
at larger lattice volumes and at other values of $\beta$.


\section*{Acknowledgments}
We thank Reinhard Alkofer, Terry Goldman, Kurt Langfeld,
Axel Maas and Dan Zwanziger
for helpful comments and suggestions
and Martin L\"uscher for sending
us the latest double-precision version of the RANLUX
random number generator.
The research of A.C. and T.M. is supported by FAPESP
(Project No.\ 00/05047-5).
A.T. thanks CNPq for financial support and the IFSC-USP
for the hospitality. 



\clearpage
\widetext

\narrowtext


%
%

\narrowtext
\begin{table}
\setcounter{table}{1}
\caption{For each coupling $\beta$ we report the value of the average
plaquette $ \langle W_{1,1} \rangle $, the string tension
$\protect\sqrt{\sigma} $ in lattice units, the lattice spacing in fm
and the smallest non-zero momentum (in MeV) for the
lattice volume $V = 140^3$.
Error bars for $ \langle W_{1,1} \rangle $ have been obtained
taking into account the integrated autocorrelation time of the HOR algorithm.
All the other error bars come from propagation of errors.}
\label{tab:volu}
\begin{tabular}{ c c c c c c c c c c c }
 $\beta$ & $\langle W_{1,1} \rangle$ & $\sqrt{\sigma}$ & $a$ (fm) &
$p_{min}$ (MeV) \\ \hline
4.2 & 0.741861(2) & 0.387(3) & 0.174(1) &  59.0(4) \\
5.0 & 0.786869(2) & 0.314(2) & 0.1407(8) & 62.9(4) \\
6.0 & 0.824780(1) & 0.254(1) & 0.1138(5) & 77.8(4) \\
\end{tabular}
\end{table}

\narrowtext
\begin{table}
\setcounter{table}{0}
\vspace*{8.65cm}
\caption{The pairs $(\beta\mbox{,} V)$ considered for the simulations,
the number of configurations, the numbers of HOR sweeps used for
thermalization and between two consecutive
configurations (used for evaluating the gluon propagator) and
the parameter $p_{sor}$ used by the SOR algorithm.}
\label{tab:runs}
\begin{tabular}{ c c c c c c }
$\beta$ & $V$ & Configurations & Thermalization &
Sweeps & $p_{sor}$ \\ \hline
4.2 &  $40^3$ & 400 & 1100 & 100 & 0.70 \\
4.2 &  $80^3$ & 200 & 2200 & 200 & 0.80 \\
4.2 & $140^3$ &  30 & 2750 & 250 & 0.88 \\ \hline
5.0 &  $40^3$ & 400 & 1320 & 120 & 0.69 \\
5.0 &  $80^3$ & 200 & 2420 & 220 & 0.80 \\
5.0 & $140^3$ &  30 & 3080 & 280 & 0.88 \\ \hline
6.0 &  $40^3$ & 400 & 1540 & 140 & 0.68 \\
6.0 &  $80^3$ & 200 & 2680 & 240 & 0.80 \\
6.0 & $140^3$ &  30 & 3300 & 300 & 0.87 
\end{tabular}
\end{table}

\narrowtext
\begin{table}
\setcounter{table}{2}
\vspace*{0.3cm}
\caption{Fit of the data using eqs.\ \protect\reff{eq:gribov}
and \protect\reff{eq:newgribov}.
In all cases $\chi^2/d.o.f.$ was of order 1.
Note that for a lattice volume $V = N^3$ we have $1 + N/2$ data points
and that all points have been used for the fits.}
\label{tab:par}
\begin{tabular}{c c c c c c c}
$\beta$ & $V$ & \multicolumn{3}{c}{Fit 1} & \multicolumn{2}{c}{Fit 2}
 \\  \cline{3-5} \cline{6-7}
        &     & $\alpha$ & $x$ & $\left|y\right|$ & $\alpha$ & $y^2$ \\ \hline
4.2 & $ 40^3$ &  0.69(2) &  0.17(2) &  0.42(1) &  0.48(4) &  0.29(1) \\ 
4.2 & $ 80^3$ &  0.66(2) &  0.17(2) &  0.36(1) &  0.46(4) &  0.24(1) \\ 
4.2 & $140^3$ &  0.61(4) &  0.19(4) &  0.35(3) &  0.38(6) &  0.25(2) \\ 
\hline
5.0 & $ 40^3$ &  0.77(3) &  0.11(2) &  0.28(2) &  0.53(6) &  0.152(8)  \\ 
5.0 & $ 80^3$ &  0.71(2) &  0.11(1) &  0.214(9) &  0.45(3) &  0.118(6)  \\ 
5.0 & $140^3$ &  0.71(3) &  0.10(2) &  0.22(2) &  0.44(6) &  0.12(1) \\ 
\hline
6.0 & $ 40^3$ &  0.86(3) &  0.069(8)  &  0.20(2) &  0.64(6) &  0.082(6)  \\ 
6.0 & $ 80^3$ &  0.84(2) &  0.032(6)  &  0.166(8) &  0.65(5) &  0.051(3)  \\ 
6.0 & $140^3$ &  0.80(2) &  0.037(8)  &  0.123(7) &  0.55(6) &  0.040(4)  \\ 
\end{tabular}
\end{table}

\end{document}